\documentclass[aps,prl,twocolumn,superscriptaddress]{revtex4-1}
\usepackage{graphicx}
\begin{document}

\title{Interplay of chiral and helical states in a Quantum Spin Hall Insulator lateral junction}
\author{M.\ R.\ Calvo}
\email[E-mail: ]{rcalvo@nanogune.eu or mreyescalvo@gmail.com}
\affiliation{Department of Physics, Stanford University, Stanford, California 94305, USA}
\affiliation{Stanford Institute for Materials and Energy Sciences, SLAC National Accelerator Laboratory, Menlo Park, California 94025, USA}
\affiliation{CIC nanoGUNE, 20018 Donostia-San Sebastian, Spain}
\affiliation{Ikerbasque, Basque Foundation for Science, 48013 Bilbao, Spain}
\author{F.\ de Juan}
\altaffiliation[Present address:]{Rudolf Peierls Centre for Theoretical Physics, Oxford University, UK}
\affiliation{Department of Physics, University of California, Berkeley, California 94720, USA}
\author{R.\ Ilan}
\affiliation{Raymond and Beverly Sackler School of Physics and Astronomy, Tel Aviv University, Tel Aviv 69978, Israel}
\affiliation{Department of Physics, University of California, Berkeley, California 94720, USA}
\author{E.\ J.\ Fox}
\affiliation{Department of Physics, Stanford University, Stanford, California 94305, USA}
\affiliation{Stanford Institute for Materials and Energy Sciences, SLAC National Accelerator Laboratory, Menlo Park, California 94025, USA}
\author{A.\ J.\ Bestwick}
\affiliation{Department of Physics, Stanford University, Stanford, California 94305, USA}
\affiliation{Stanford Institute for Materials and Energy Sciences, SLAC National Accelerator Laboratory, Menlo Park, California 94025, USA}
\author{M.\ M\"uhlbauer}
\affiliation{Physikalisches Institut (EP3) and R\"ontgen Center for Complex Material Systems,
Universit\"at W\"urzburg, Am Hubland, 97074 W\"urzburg, Germany}
\author{J.\ Wang}
\affiliation{Department of Physics, Stanford University, Stanford, California 94305, USA}
\affiliation{Stanford Institute for Materials and Energy Sciences, SLAC National Accelerator Laboratory, Menlo Park, California 94025, USA}
\affiliation{State Key Laboratory of Surface Physics and Department of Physics, Fudan University, Shanghai 200433, China}
\author{C.\ Ames}
\affiliation{Physikalisches Institut (EP3) and R\"ontgen Center for Complex Material Systems,
Universit\"at W\"urzburg, Am Hubland, 97074 W\"urzburg, Germany}
\author{P.\ Leubner}
\affiliation{Physikalisches Institut (EP3) and R\"ontgen Center for Complex Material Systems,
Universit\"at W\"urzburg, Am Hubland, 97074 W\"urzburg, Germany}
\author{C.\ Br\"une}
\affiliation{Physikalisches Institut (EP3) and R\"ontgen Center for Complex Material Systems,
Universit\"at W\"urzburg, Am Hubland, 97074 W\"urzburg, Germany}
\author{S.\ C.\ Zhang}
\affiliation{Department of Physics, Stanford University, Stanford, California 94305, USA}
\affiliation{Stanford Institute for Materials and Energy Sciences, SLAC National Accelerator Laboratory, Menlo Park, California 94025, USA}
\author{H.\ Buhmann}
\affiliation{Physikalisches Institut (EP3) and R\"ontgen Center for Complex Material Systems,
Universit\"at W\"urzburg, Am Hubland, 97074 W\"urzburg, Germany}
\author{L.\ W.\ Molenkamp}
\affiliation{Physikalisches Institut (EP3) and R\"ontgen Center for Complex Material Systems,
Universit\"at W\"urzburg, Am Hubland, 97074 W\"urzburg, Germany}
\author{D.\ Goldhaber-Gordon}
\email[E-mail: ]{goldhaber-gordon@stanford.edu}
\affiliation{Department of Physics, Stanford University, Stanford, California 94305, USA}
\affiliation{Stanford Institute for Materials and Energy Sciences, SLAC National Accelerator Laboratory, Menlo Park, California 94025, USA}

\date{\today}

\begin{abstract}
%
We study the electronic transport across an electrostatically-gated lateral junction in a HgTe quantum well, a canonical 2D topological insulator, with and without applied magnetic field. We control carrier density inside and outside a junction region independently and hence tune the number and nature of 1D edge modes propagating in each of those regions. Outside the bulk gap, magnetic field drives the system to the quantum Hall regime, and chiral states propagate at the edge. In this regime, we observe fractional plateaus which reflect the equilibration between 1D chiral modes across the junction. 
As carrier density approaches zero in the central region and at moderate fields, we observe oscillations in resistance that we attribute to Fabry-Perot interference
in the helical states, enabled by the broken time reversal symmetry. At higher fields, those oscillations disappear, in agreement with the expected absence of helical states when band inversion is lifted.


\end{abstract}

\maketitle


Above a certain critical thickness, the 2-dimensional electron gas (2DEG) of a HgTe quantum well presents an inverted band structure 
characteristic of a 2D topological insulator (2D-TI) \cite{qi2011topological,hasan2010colloquium}. At the edge of the topological insulator, quantum spin Hall (QSH) helical states propagate \cite{kane2005quantum,bernevig2006quantum}. 
When the Fermi level lies in the bulk gap of a 2D-TI, conduction is dominated by those edge states \cite{konig2007quantum,konig2008quantum,konig2007quantum,konig2008quantum} and is in principle protected by time-reversal symmetry (TRS) against single-electron backscattering processes.
The application of magnetic field is expected to lift such 
protection. Nonetheless, band inversion 
and counterpropagating QSH-like edge states are predicted to persist up to a critical magnetic field $B_{\rm c}$. Above this field, band inversion should disappear, leaving a 2D band structure identical to that of a \textit{topologically trivial} \footnote{The electronic structure of a 2DEG in the Quantum Hall regime is also topologically nontrivial. Here, however, we use trivial to describe only the lowest energy bands characterized by the Z2 topological invariant.} semiconductor in the Quantum Hall (QH) regime \cite{konig2007quantum,konig2008quantum,scharf2012magnetic}. 

Previous experiments on HgTe quantum wells in this thickness range show that the resistance in the bulk gap increases in the presence of moderate magnetic fields~\cite{konig2008quantum,gusev2013linear,maciejko2010magnetoconductance} as predicted \cite{tkachov2010ballistic,scharf2012magnetic}, but our understanding of the evolution of edge conduction with magnetic field is incomplete.
For the related problem of assigning quantum numbers to different chiral quantum Hall (QH) modes, a fruitful approach has been to study scattering between those modes by measuring transport through junctions between regions of different carrier density.
This approach has been widely applied in GaAs quantum wells (for a review see Ref.~\cite{haug1993edge}) and more recently in the Dirac 2DEG of graphene \cite{williams2007quantum, ozyilmaz2007electronic,ki2009quantum,amet2014selective}. Thus, to characterize helical modes under broken TRS, studying their interplay with quantum Hall chiral states could be a promising strategy.


In this work, we explore electronic transmission across a lateral heterojunction fabricated on a HgTe quantum well with inverted band structure. 
Above a critical field, our results are consistent with expectations for equilibration of QH edge modes.  Results are similar below the critical field for high carrier densities, but clearly differ when the junction is tuned through zero density. There, we first observe how the maximum of resistance associated with the bulk gap narrows and shifts towards lower values of carrier density. We find this to be a consequence of the remaining band inversion and the existence of helical edge states. 
Over the density regime corresponding to the peak shift, the resistance of our device presents oscillations which we attribute to Fabry-Perot interference of helical states enabled by the lifting of TRS protection. 

\begin{figure}
\includegraphics[width=\linewidth]{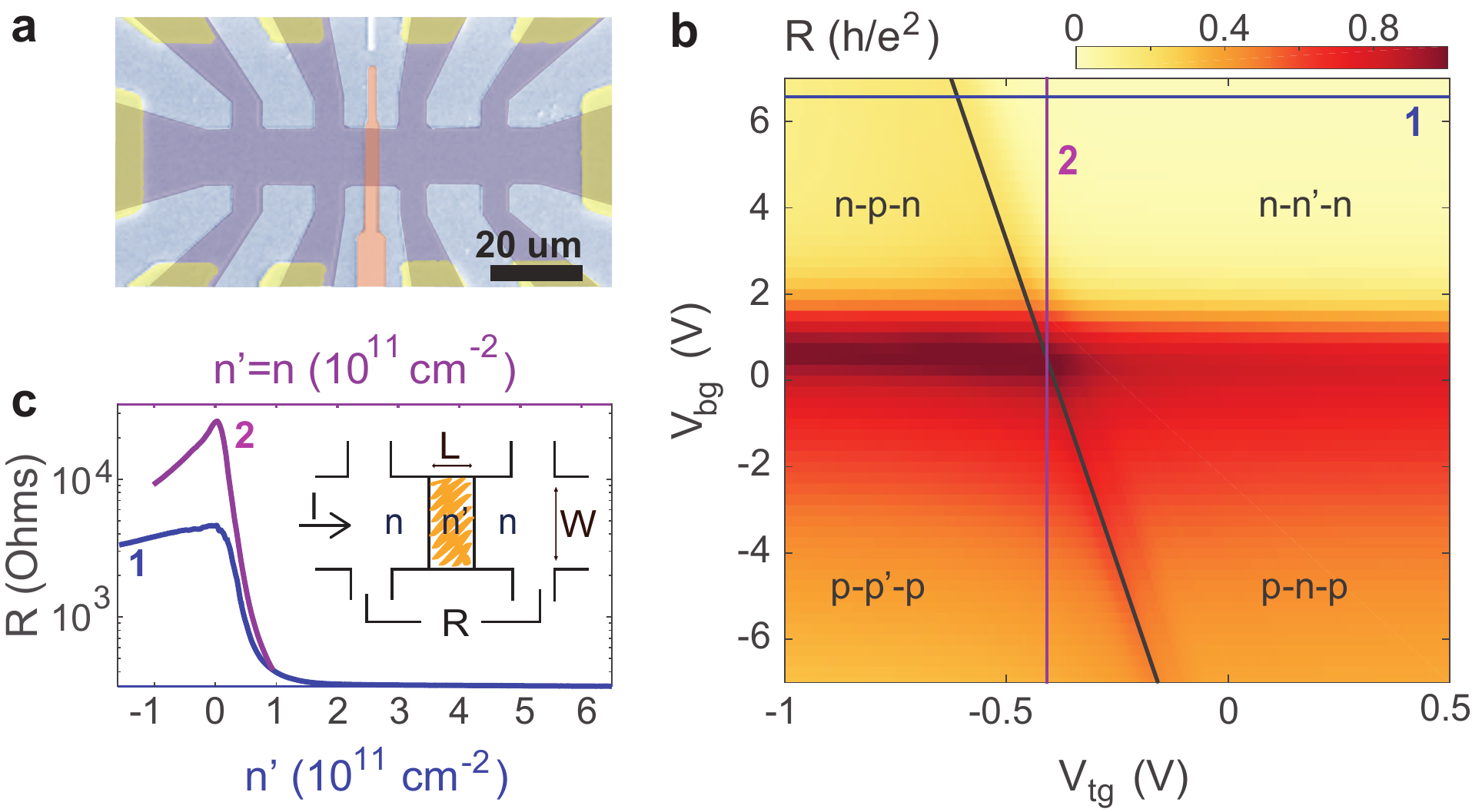}
\caption{\label{Fig1} (a) Electron micrograph of Hall bar device with narrow top gate. The geometry of the device is sketched in the inset of panel (c). The width ($W$) of the Hall bar is 10 $\mu$m and the physical length ($L$) of the top gate is 2 $\mu$m. Carrier density in the outer and central regions of the junction are denoted by $n$ and $n'$ respectively.  (b) 4-terminal resistance ($R$) across top-gated region measured 
as a function of top gate and back gate voltages at zero applied magnetic field. The diagonal black line marks the estimated position of $n'=0$. (c) Selected linecuts extracted from panel (b): Linecut 1 (blue) shows the evolution of $R$ as a function only of $V_{\rm tg}$, that is, as a function of $n'$ while the outer region is kept at constant $n=10^{11}$ cm$^{-2}$. Linecut 2 shows $R$ as function solely of $V_{\rm bg}$, which changes the density of the whole device, for a value of $V_{\rm tg}=V_{\rm tg0}$ such that the device density is homogeneous, that is, $n'\simeq n$.}
\end{figure}

Fig.~\ref{Fig1}(a) presents the geometry of our device. A Hall bar mesa is defined following the method described in Refs.~\cite{ma2015unexpected,SM} on a HgTe quantum well with inverted band structure. The 2DEG is formed in a quantum well epitaxially grown over a conductive substrate \cite{Leubner_2016,SM}, allowing for the application of an overall back-gate voltage. 
A narrow top gate electrode is placed at the center of the device, defining two regions with separately-tunable density (inset of Fig.~\ref{Fig1}(c)). \textit{Central region} refers to the area covered by the top-gate electrode and \textit{outer region} to its surroundings. The carrier density in the outer region $n$ can be tuned by the applied back-gate voltage $V_{bg}$, while the density $n'$ in the central region depends on both back- and top-gate ($V_{tg}$) voltages. Both $n$ and $n'$ can be estimated using a simple capacitor model \cite{SM}.

The evolution of the four-terminal resistance $R$ measured across the junction at 2.1 K (inset of Fig.~\ref{Fig1}(c)) as a function of both $V_{bg}$ and $V_{tg}$ at zero applied magnetic field is shown in Fig.~\ref{Fig1}(b). As previously reported \cite{konig2007quantum}, the resistance presents a finite maximum when the chemical potential lies in the bulk gap and conduction is dominated by the QSH edge states. As a function of spatially-uniform carrier density, four-terminal resistance (Curve 2 in Fig.~\ref{Fig1}(c)) shows a maximum value higher than $h/2e^2$, the value associated with the ballistic Quantum Spin Hall regime. This is expected: the edge mean free path in similar heterostructures has been reported to be a few microns \cite{konig2007quantum}, substantially less than the 
edge length between contacts in the present geometry, so backscattering should result in increased resistance. In contrast, as a function only of density in the central region $n'$ (Curve 1 in Fig.~\ref{Fig1}(c)) the maximum resistance is lower than $h/2e^2$, suggesting the presence of bulk conduction in parallel to the QSH edge states. A detailed look at the data (Fig.~S2 of \cite{SM}) reveals oscillations in the resistance which likely arise from the Fabry-Perot like interference of
bulk conduction paths
(for a detailed analysis see \cite{SM}).

The locations of resistance maxima in the $(V_{\rm tg},V_{\rm bg})$ parameter space (Fig.~\ref{Fig1}(a)) fall along two lines: a horizontal line around $V_{\rm bg}=V_{\rm bg0}=0$ V representing zero density in the outer region, and a diagonal line representing zero density in the central region of the junction ($n'=0$) (see \cite{SM}).
The two lines define four quadrants of electron and hole densities in the central and outer regions of the junction, labeled in Fig.~\ref{Fig1}(b).



\begin{figure}
\includegraphics[width=\linewidth]{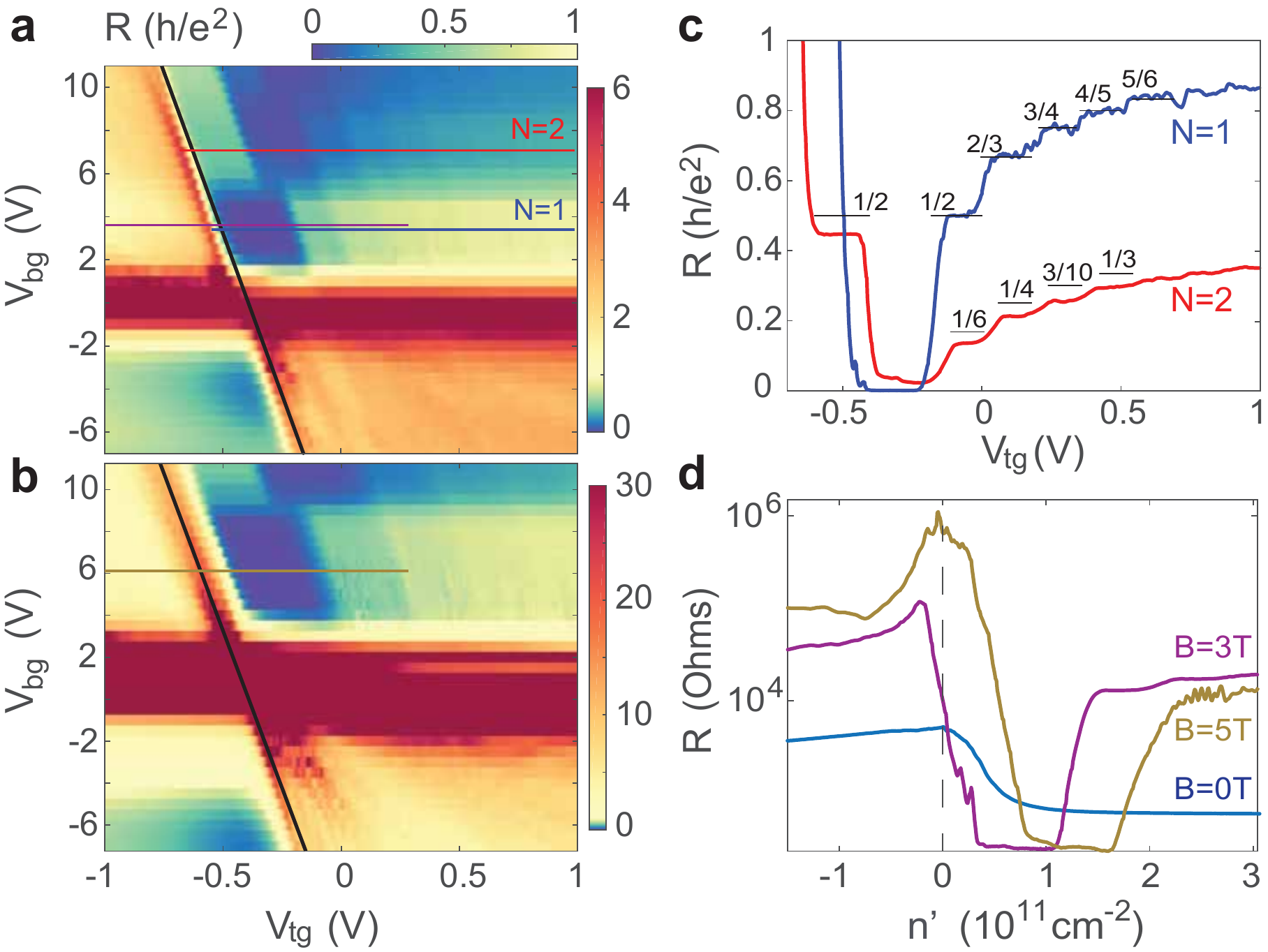}
\caption{\label{Fig2} 
(a),(b) 2D maps of resistance obtained at B = 3 and 5 T respectively. The color scale has been chosen to enhance the contrast in the $n-n'-n$ region between 0 and 1 $h/e^2$. The fractional resistance values match predictions for electron transmission from $N$ QH edge modes in the outer regions of the junction into $N'$ modes in the central region in the presence of edge mode equilibration.
(c) Horizontal linecuts from panel (a), at $V_{\rm bg}=3.5$ V (blue) and 7 V (red), corresponding to $N=1$ and $N=2$. $N'$ is tuned by the top-gate voltage $V_{\rm tg}$. (d) Four-terminal resistance $R$ as a function of density in the junction for $B=0$ (blue), $B=3$ (magenta), $B=5$ T (yellow) for density in the outer region $n=5\times 10^{10}$, $5\times 10^{10}$, $8\times 10^{10}$ cm$^{-2}$ respectively. For the cases with finite applied field, those carrier densities correspond to $N=1$.}
\end{figure}

At finite fields, four-terminal resistance $R$ in the $n$-$n'$-$n$ quadrant shows a sharp tiled pattern of fractional resistance values ranging from 0 to $h/e^2$ (Fig.~\ref{Fig2}(a,b), $B=3$ and 5 T respectively.)
The Landauer-B\"{u}ttiker formalism 
describes those fractional values as the result of full equilibration between co-propagating edge states in the junction (cf.~\cite{haug1993edge,williams2009quantum,ozyilmaz2007electronic}.) In the unipolar regime and in a 4-terminal configuration, the predicted resistance across the junction is given by
\begin{equation}
R=\frac{h}{e^2}\frac{N-N'}{NN'}
\label{lb}
\end{equation}
where $N$ and $N'$ are the number of quantum Hall states propagating in the outer and central regions of the junction respectively. The bottom row of tiles in Fig.~\ref{Fig2}(a,b)
corresponds to $N=1$, with $N'$ increasing from left to right. Our data show a good agreement with the fractional plateaus at $R=0,1/2,2/3,3/4,...$ expected for $N'=1,2,3,4,...$ ($N=1$ linecut in Fig.~\ref{Fig2}(c)). Similarly, a second row of tiles appears for $N=2$, and in the corresponding linecut of Fig.~\ref{Fig2}(a), plateaus of resistance can be observed near the expected values for $N=2$ paired with a range of $N'$, although deviations from the ideal behavior are larger here than for $N=1$. Similarly, plateaus are observed near but not exactly at the expected values for the $p$-$p'$-$p$ and the $n$-$p$-$n$ quadrants (see~\cite{SM} for details.)

These results highlight the role of the strong spin-orbit interaction and inversion symmetry breaking in HgTe. If the $s_z$ component of spin were conserved, transmission would then be spin-selective and only those states with same spin polarization would equilibrate with each other, leading to fewer plateaus in resistance $R$. In contrast, our data suggest that full equilibration occurs for all possible values of $N$ and $N'$. Inversion symmetry breaking provides a mechanism spin mixing that allows this to happen~\cite{muller1992equilibration}. Adapting the theory in Ref.~\onlinecite{Khaetskii92}, we estimate that the equilibration length for our system is around 2 $\rm{\mu m}$ at 2 T, which is indeed smaller than the junction width.

In the $n$-$n'$-$n$ quartet, the tiled structure of fractional resistance values associated to a given pair of values ($N$,$N'$)  is similar at both 3 and 5 T (Fig.~\ref{Fig2}(a) and (b) respectively). 
A contrasting behavior emerges around zero density. First, the zero density $n'=0$ line 
determined from zero magnetic field data in Fig.~\ref{Fig1}(b) is overlaid for reference on Fig.~\ref{Fig2}(a) and (b). The maximum of resistance at $B=3$ T is clearly shifted towards lower values of $V_{\rm tg}$ with respect to that line.  Remarkably, it returns to the original position at  $B= 5$ T (Fig.~\ref{Fig2}(b)).
This effect can be also observed in the horizontal linecuts taken from the corresponding 2D resistance plots at similar outer densities $n$ for 0, 3 and 5 T (Fig.~\ref{Fig2}(d)).  
Furthermore, at 3T and near zero density, in the range of voltages where the resistance peak was found at zero field, this plot now shows strong oscillations in the resistance. 

\begin{figure}
\includegraphics[width=\linewidth]{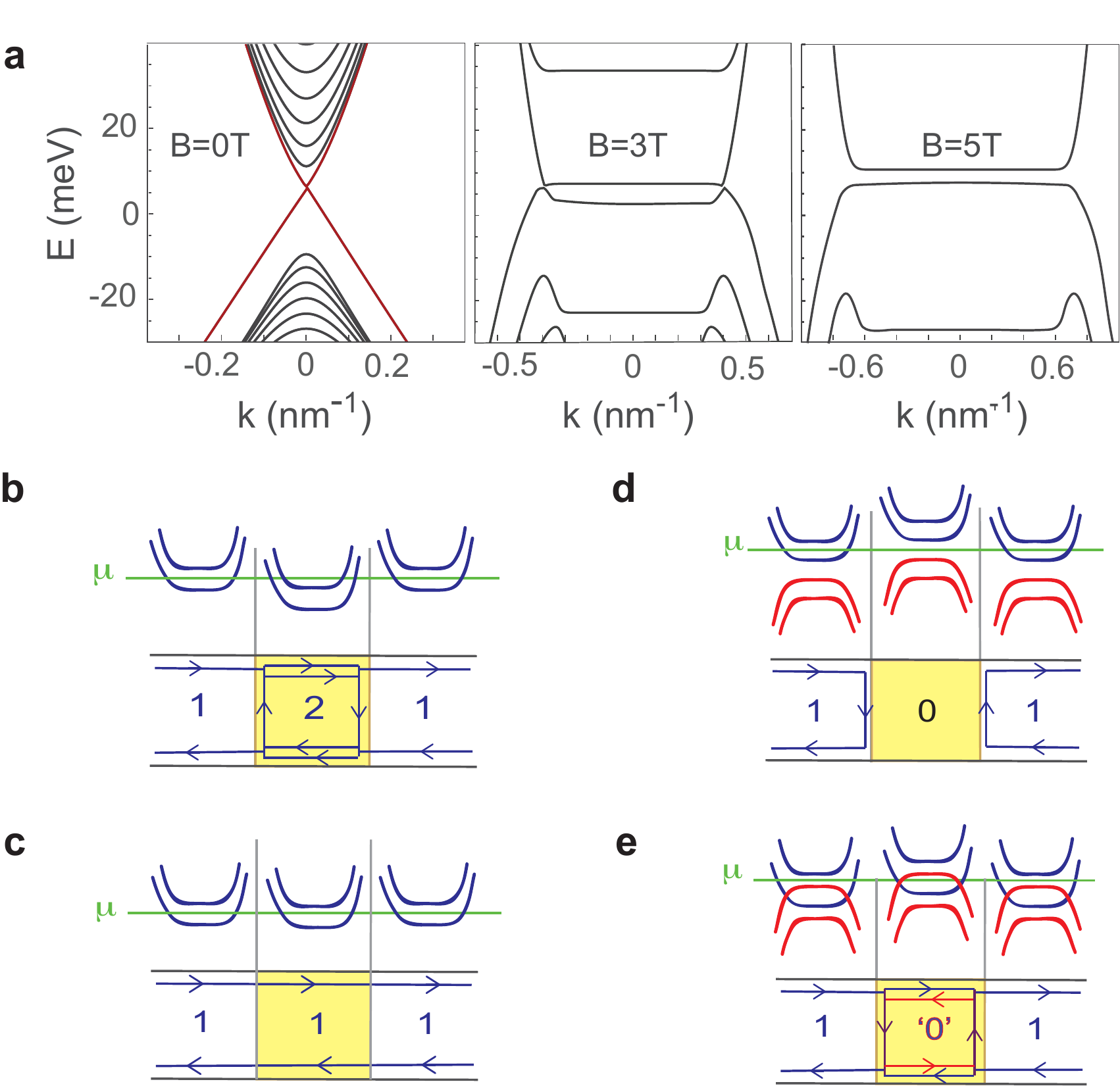}
\caption{\label{Fig3} 
(a) Computed band structure of a strained 7.5 nm HgTe quantum well at 0, 3 and 5 T. The critical field $B_{\rm c} =3.8$ T in the model. See~\cite{SM} for calculation details. (b),(c) Sketch of band structure and edge states in different scenarios at 3 and 5 T where the outer region hosts one chiral edge state and the inner region hosts either two or one, respectively. (d),(e) Similar sketches for the case $N=1$ and $N'=0$ for 5 and 3 T. (d) Above $B_{\rm c}$ a broad gap opens between the electron and hole Landau levels so no states propagate at the junction. (e) Below $B_{\rm c}$, band inversion remains and at $\nu=0$ a helical mode propagates in the junction.}
\end{figure}

To explicate these results,  we present calculations of band structure below and above the critical field, (details in \cite{SM}) for the magnetic fields considered in the experiment: $B =0,3,5$ T (Fig.~\ref{Fig3}(a)).
At zero field, when the Fermi energy lies in the bulk gap, we find the usual counterpropagating helical edge states. At both $B=3$ and $5$~T conduction and valence bands turn into a set of discrete Landau levels (LLs). 
One chiral Quantum Hall state propagates at the edge for each filled Landau level in the bulk, so the total number of modes $N$ is given by the integer part of the filling factor $\nu$. 
For fields $B < B_{\rm c}$ such as $B = 3$ T, the lowest order hole-like and electron-like Landau levels are inverted in the bulk and cross near the edge, so when the Fermi level is in the bulk gap there are counter-propagating helical edge states (below we refer to this regime as $\nu=0$). By 5 T, which is above $B_{\rm c}$, band inversion has disappeared, and the band structure resembles that of a \textit{trivial} semiconductor, with a gap between electron and hole Landau levels. 

In the junction geometry considered in our experiment, 
at finite field the electronic transmission across the device will
result from the matching of edge modes corresponding to different fillings in the central and outer regions of the junction ($\nu'$ and $\nu$ respectively).

When the central region has $\nu' \neq 0$, the edge mode structure is the usual one observed in quantum Hall experiments with standard 2DEGs 
(see Figs.~\ref{Fig3}(b,c)) 
and the resulting resistance 
is given by a Landauer-B\"uttiker expression (Eq.~\ref{lb}). 
 When $\nu'=0$ however, the situation changes drastically depending on whether $B$ is larger or smaller than $B_{\rm c}$. Above the critical field, the Fermi level always lies in a bulk gap with no edge modes, and incoming modes are always reflected, as illustrated in Fig.~\ref{Fig3}(d) for the case 1-0-1. Below the critical field, in contrast, band inversion implies the presence in the inner region of two QSH-like edge states with opposite chiralities (Fig.~\ref{Fig3}(e)). Edge modes cannot simply terminate, so a mode must also propagate along the 1-0 and 0-1 interfaces. 

\begin{figure}
\includegraphics[width=\linewidth]{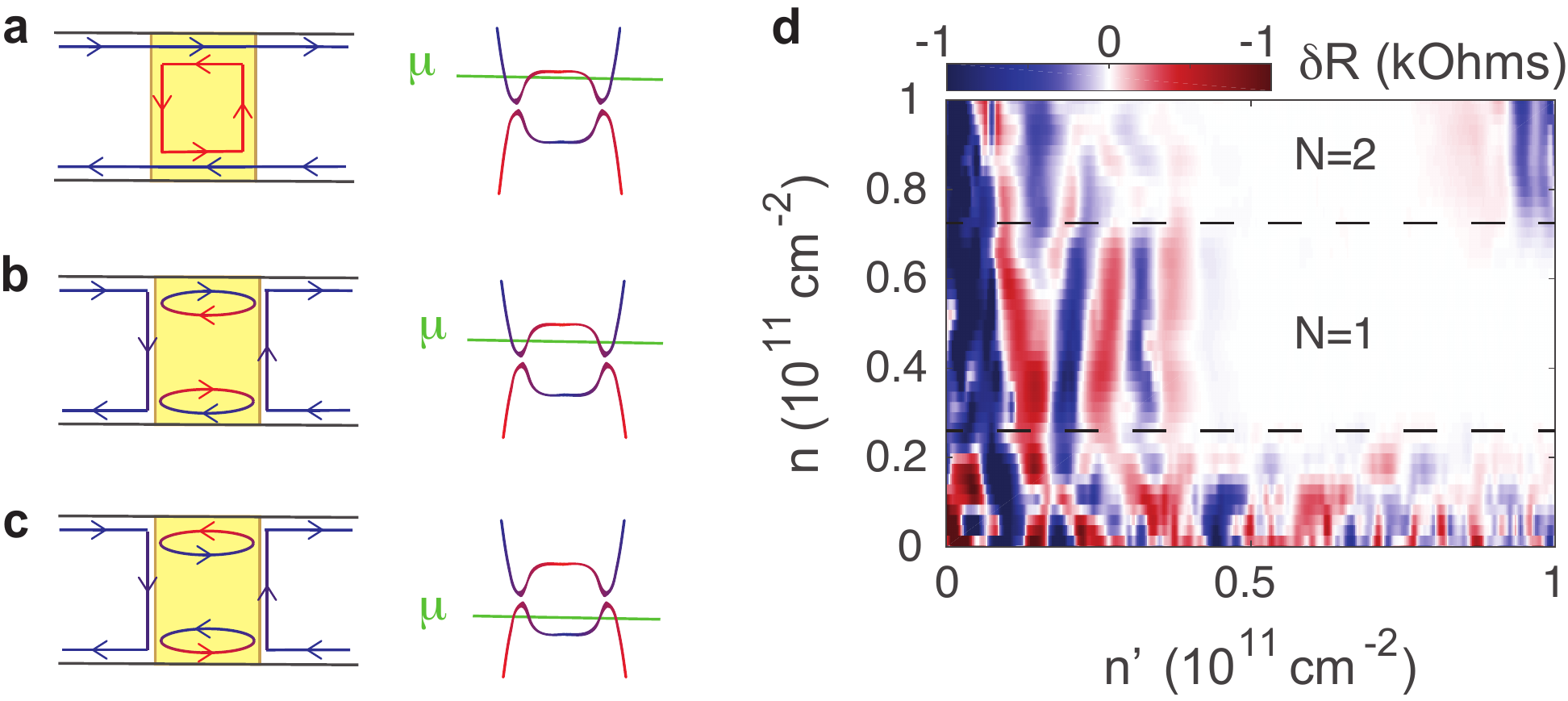}
 \caption{\label{Fig4} 
(a),(b),(c) Possible scenarios for edge state matching in the 1-0-1 situation.  (d) Zoom in the $\delta R$ oscillations (obtained by removal of a smoothed background resistance \cite{SM}) as a function of 2D density in the central and outer regions of the junction, $n, n'$, at applied field $B=3$ T. Areas with different numbers of quantum hall modes ($N=1$ and $N=2$),  i.e. integer filling factor, in the outer region are separated by dotted lines.}
\end{figure}

We believe that the matching of chiral to helical edge states in the 1-0-1 scenario is the origin of the resistance oscillations and the shift in the position of the resistance maximum we observe at 3 T 
(Fig.~\ref{Fig2}(d)).
To understand this, we first note that since TRS is broken at finite field, the crossing of QSH edge modes when $B<B_{\rm c}$ is only protected in the presence of extra symmetries such as mirror symmetry.
In the experiment, such symmetries are absent, so there should always be a small minigap.
(Figs.~\ref{Fig4}a-c). 
The location of this edge minigap within the bulk gap depends on details such as the potential at the edge. Therefore the edge and bulk charge neutrality points do not necessarily occur for the same value of gate voltage, and accordingly the center of the resistance maximum originating from the minigap at finite field does not necessarily align with the center of the bulk gap. The observed gate voltage position shift and narrowing of the resistance maximum at $3$ T compared to zero or $5$ T (see Fig.~\ref{Fig2}(d)) is consistent with an origin in the edge state minigap. 

In the 1-0-1 configuration, when the incoming chiral edge mode from the outer region reaches the junction it can scatter into two possible outgoing modes: the co-propagating helical edge mode or the chiral mode parallel to the junction. When the chemical potential in the central region is very close to the bottom of the lowest Landau Level, the incoming edge mode is almost perfectly matched to the copropagating helical one, while the counterpropagating edge mode forms a loop spanning the whole junction (Fig.~\ref{Fig4}(a)). This must be so because the counterpropagating mode has smaller momentum and therefore is located farther from the edge. Transport in this scenario is almost equivalent to the 1-1-1 situation, seen in the experimental data as an extension of the $R=0$ plateau to lower densities (Fig.~\ref{Fig2}(a)).


As the chemical potential approaches the crossing of the helical edge modes, the chiral mode connects to the one parallel to the junction, while the helical modes form a loop at either edge (Fig.~\ref{Fig4}(b)).
These loops should disappear at the minigap, and reappear with opposite orientation below it (Fig.~\ref{Fig4}(c)). The existence of these loops, allowed because the protection from backscattering is lifted by $B$, implies that coherent transport should be affected by multiple reflections at the interfaces. This effect should manifest in Fabry-Perot type oscillations as a function of chemical potential, because the accumulated phase $\delta = k L$ depends smoothly on chemical potential. This explains the oscillations observed at 3 T in Fig.~\ref{Fig2}(d), in the density range assigned to the bulk gap and adjacent to the edge minigap,
and their disappearance beyond $B_{\rm c}$ (i.e. at 5 T) where no helical modes exist. 

Resistance oscillations are also present at zero field in a similar density regime -- these we associate with bulk states. However, at 3T, oscillations present an amplitude about an order of magnitude higher than their zero field counterparts. Moreover, our data (Fig.~\ref{Fig2}(c) and Fig. S4(d)) indicate that the bulk states causing the oscillations at zero field must be fully localized at 3T (see \cite{SM}), further suggesting that oscillations at 3T have their origin in edge rather than bulk states.

Furthermore, oscillations are periodic in $n'$, with no substantial dependence on $n$ (Fig. 4(d)). This is consistent with our interpretation: changing $n'$ will change the edge momentum of the loop modes and hence the phase, while $n$ only determines the momentum of the incoming modes, which should have no effect on the phase of the oscillations. Fig.~\ref{Fig4}(d) also shows that oscillations are present for $n$ values corresponding to $N$=1, but they disappear when approaching $N=2$. This is consistent with the $N$=2 Quantum Hall chiral edge modes not being fully established at $B=3$ T, as already seen in the imperfectly-quantized equilibration plateaus in Fig.~\ref{Fig2}(c). 

Taken together, the amplitude, position and field- and carrier density-dependence of resistance oscillations support our interpretation of their origin in the constructive interference of helical edge states.

Finally, assuming a Fabry-Perot scenario yields some quantitative estimates for system parameters. Based on bulk 2D Fabry-Perot oscillations at $B=0$ we estimate the effective length $L^*$ of the central region to be $0.6$ $\mu$m (see section 4 of \cite{SM} for details). This length need not be the same as the physical top-gate length, due to the smooth shape of the gate-induced potential. Given that we observe coherent FP oscillations from the QSH edge states as well, a lower bound on the edge localization length can be set at $L^*$. We expect that at larger channel lengths or with higher disorder, the QSH loop responsible for interference will break up into more loops and coherence will gradually be lost, in a way similar to Ref. ~\onlinecite{tkachov2010ballistic}. We may also estimate the 1D edge mode carrier density: given the above value of $L^*$ we infer $C_{\rm 1D}=\delta(n_{1D})/\delta(V_{\rm tg})\simeq 1.4\times 10^6$ cm$^{-1}$V$^{-1}$ (see section 6 of \cite{SM} for details). 

While our results below critical field are compatible with those in Ref.~\cite{gusev2013quantum}, we present here evidences for physical scenarios that were not accessible in that work. More specifically, the dual-gate configuration of our device allows us to perform a detailed study of the equilibration of QH states in HgTe QWs and to infer the role played here by spin-orbit interaction. More importantly, we present one of very few evidences for a transition between QSH and QH regimes in a 2D-TI. Finally,
we observe signatures of coherent interference on helical states, likely due to the geometry of our junction. 
Our results suggest that valuable information about the QSH state under broken TRS can be inferred from the electronic transmission across a QH-QSH-QH heterojunction.

While the present work was under review, a related theoretical work by Nanclares et al. \cite{nanclares} has been published.

  

\begin{acknowledgments}
The work at Stanford was supported by the
Department of Energy, Office of Science, Basic Energy Sciences, Materials Sciences and Engineering Division, under contract DE-AC02-76SF00515 to D.G.-G. and S.-C.Z.
The Center for Probing the Nanoscale, an NSF NSEC under grant PHY-0830228 to D.G.-G. supported early stages of the project. The European Union under the project FP7-PEOPLE-2010-274769 supported M.R.C.'s stay at Stanford. We also acknowledge support from the National Thousand-Young-Talents Program to J.W. and funding from an AFOSR MURI (F.J.) and from the DARPA FENA program (R.I.).
The W\"urzburg group acknowledges additional financial support from the German Research Foundation (The Leibniz Program, Sonderforschungsbereich 1170 ‘Tocotronics’ and Schwerpunktprogramm 1666), the EU ERC-AG program (Project 3-TOP), the Elitenetzwerk Bayern IDK ‘Topologische Isolatoren’ and the Helmoltz Foundation (VITI).

\end{acknowledgments}

\bibliography{hgte.bib,heterojunctions.bib}
\end{document}